\documentclass{webofc}
\usepackage[varg]{txfonts}   
\usepackage{bm}
\newcommand{\beq}{\begin{equation}}
\newcommand{\eeq}{\end{equation}}
\begin{document}
\title{Relativistic Brueckner-Hartree-Fock Theory in Infinite Nuclear Matter}
%
%

\author{\firstname{Peter} \lastname{Ring}\inst{1,2}\fnsep\thanks{\email{peter.ring@tum.de}} \and
        \firstname{Sibo} \lastname{Wang}\inst{2}\and
        \firstname{Qiang} \lastname{Zhao}\inst{2}\and
        \firstname{Jie} \lastname{Meng}\inst{2}
}

\institute{
Fakult\"at f\"ur Physik, Technische Universit\"at M\"unchen, D-85747 Garching, Germany
\and
State Key Laboratory of Nuclear Physics and Technology, School of Physics, Peking University, Beijing 100871, China
}

\abstract{
On the way of a microscopic derivation of covariant density functionals, the first complete solution of the relativistic Brueckner-Hartree-Fock (RBHF) equations is presented for symmetric nuclear matter. In most of the earlier investigations, the $G$-matrix is calculated only in the space of positive energy solutions. On the other side,
for the solution of the relativistic Hartree-Fock (RHF) equations, also the elements of this matrix connecting positive and negative energy solutions are required. So far, in the literature, these matrix elements are derived in various approximations. We discuss solutions of the Thompson equation for the full Dirac space and compare the resulting equation of state with those of earlier attempts in this direction.
}
\maketitle
\section{Introduction}
\label{intro}

In recent years, microscopic theories of the nuclear many-body problem, starting with the bare nucleon-nucleon
forces, showed considerable progress in describing the properties of light and even specific medium-heavy nuclei~\cite{Baldo2007_JPG734-R243,Barrett2013_PPNP69-131,Hagen2014_RPP77-096302,Carlson2015_RMP87-1067,
Hegert2016_PR621-165,SHEN-SH2019_PPNP109-103713}. However, the largest part of the nuclear chart is still only accessible in energy density functional (EDF) theories~\cite{Bender2003_RMP75-121} which are based on the mean-field concept. In Coulombic systems, where density functional theory has been originally introduced by Kohn and Sham~\cite{Kohn1965_PR137-A1697,Kohn1965_PR140-A1133} such functionals can, nowadays, be derived, in a fully microscopic way, from the bare Coulomb force~\cite{Perdew2003_LNP620-269}, but in nuclei the situation is much more complex. First, the extreme strength of the bare nucleon-nucleon force with a strongly repulsive core at short distances and an attractive Yukawa tail caused by pion exchange forbids any mean-field approximation. On the other side, the existence of shell effects~\cite{Haxel1949_PR75-1766,GoeppertMayer1950_PR78-16} and the success of the phenomenological mean-field models~\cite{Nilsson1955_MFMDVS29-16} show that nuclei can be well described within an averaged field. Therefore, already in the fifties the concept of an effective force has been introduced by Brueckner~\cite{Brueckner1955_PR97-1353,Day1967_RMP39-719}: nucleons in the medium do not feel the strong bare force. The Pauli principle forbids the scattering into occupied orbits and reduces this force. Taking this effect into account in the scattering theory, Brueckner replaced the Lippman-Schwinger equation~\cite{Lippmann1950_PR79-469} for the free scattering by the Bethe-Goldstone equation~\cite{Bethe1957_PRSA238-551} for the scattering in the nuclear medium and introduced the $G$-matrix that depends on the density. In the fifties, neither were the bare nucleon forces known well enough, nor the computer facilities were strong enough to solve the corresponding self-consistent equations successfully. In the sixties and seventies, when the bare forces were better known and when computers became available to solve the Brueckner-Hartree-Fock (BHF) equations, it turned out that this concept failed in describing the proper saturation properties of nuclear matter~\cite{Coester1970_PRC1-769}. On the other side, nuclear density functionals with phenomenological, effective, density-dependent, Skyrme \cite{Vautherin1972_PRC5-626} and Gogny \cite{Decharge1980_PRC21-1568} interactions turned out to be very successful. Covariant density functional theories (CDFTs)~\cite{Walecka1974_APNY83-491,Boguta1977_NPA292-413,Serot1986_ANP16-1,Reinhard1989_RPP52-439,
Ring1996_PPNP37-193,Meng2006_PPNP57-470} are of particular interest because Lorentz invariance is one of the underlying symmetries of QCD. This symmetry not only allows to describe the spin-orbit coupling, which has an essential influence on the underlying single-particle structure, but it also puts stringent restrictions on the number of parameters in the corresponding functionals without reducing the quality of the agreement with experimental data~\cite{Meng2016}.

Brueckner theory is only an approximation, but other methods, for instance variational methods, also failed~\cite{Pandharipande1979_RMP51-821}. In this situation, two new concepts have been introduced in the following years: (a) The concept of deriving the properties of finite nuclei only from bare two-nucleon forces has been given up and three-body forces have been introduced in various microscopic theories~\cite{Fujita1957_PTP17-360,ZUO-W2002_NPA706-418,Hammer2013_RMP85-197}. (b) Lorentz invariance was taken into account.
This leads to relativistic Brueckner theory. In fact, the Brooklyn group~\cite{Anastasio1980_PRL45-2096} could show that an approximate solution of the relativistic Brueckner-Hartree-Fock equations can reproduce the saturation properties of nuclear matter rather well. Over the years many groups worked in this direction~\cite{Horowitz1987_NPA464-613,Poschenrieder1988_PRC38-471,Brockmann1990_PRC42-1965,deJong1998_PRC58-890,Gross-Boelting1999_NPA648-105,
Schiller2001_EPJA11-15,MA-ZY2002_PRC66-024321,Fuchs2004_LNP641-111,Katayama2015_PLB747-43} and solved the relativistic BHF problem in various approximations.

In this contribution we discuss the relativistic Brueckner-Hartree-Fock problem in detail and show results of new
full solutions, where the positive energy states (PES) and the negative energy states (NES) of the relativistic HF problem are fully taken into account. We also compare with approximate treatments of the NES.

\section{Relativistic Brueckner-Hartree-Fock Theory}
\label{sec-1}

The full solution of the Brueckner-Hartree-Fock equations is an iterative process. For each density characterized by the Fermi momentum $k_F$ an effective interaction, the $G$-matrix, is determined by the solution of the Bethe-Goldstone equation~\cite{Bethe1957_PRSA238-551}:
\beq\label{20190619-eq1}
  \hat{G}(W) = \hat{V}  + \hat{V}\frac{\hat{Q}}{W-\hat{H}_{HF}}\hat{G}(W).
\eeq
Here $W$ is the starting energy. $\hat{Q}$ is the Pauli operator, which allows only scattering processes into intermediate states above the Fermi surface in the nuclear system, and $\hat{H}_{HF}$ is the Hartree-Fock single particle operator for the motion of the two particles in the intermediate states.
\beq\label{HF}
  \hat{H}_{HF} = \hat{T} + \hat{U},
\eeq
where the the matrix elements of the single-particle potential are given by
\beq\label{HF}
  \langle n|\hat{U}|n'\rangle=\sum_{k<k_F} \langle nk|\hat{G}|n'k\rangle.
\eeq
The index $k$ runs over all the states in the Fermi sea below the Fermi level.

In the relativistic case, the free scattering process is usually described by the Thompson equation~\cite{Thompson1970_PRD1-110}, one of the relativistic three-dimensional reductions of the Bethe-Salpeter equation~\cite{Salpeter1951_PR84-1232}. In particular, this equation has been used for the adjustment of the relativistic bare nucleon-nucleon potential Bonn A~\cite{Machleidt1989_ANP19-189} to the experimental phase shifts for scattering processes of two particles with positive energies.

In RBHF theory, i.e. for the scattering process in infinite nuclear matter, the Thompson equation has the form
\begin{equation}\label{eq-4}
  G(\bm{q}',\bm{q}|\bm{P},W)
  = V(\bm{q}',\bm{q}|\bm{P})
  + \int \frac{d^3k}{(2\pi)^3}  V(\bm{q}',\bm{k}|\bm{P})
    \frac{Q(\bm{k},\bm{P})}{W-E_{ \bm{P}+\bm{k}}-E_{ \bm{P}-\bm{k}}}G(\bm{k},\bm{q}|\bm{P},W),
\end{equation}
where
$\bm{P}=\frac{1}{2}({\bm k}_1+{\bm k}_2)$ is the center-of-mass (c.m.) and $\bm{k}=\frac{1}{2}({\bm k}_1-{\bm k}_2)$ is the relative momentum of the two nucleons in nuclear matter with momenta ${\bm k}_1$ and ${\bm k}_2$. $\bm{q}, \bm{k}$, and $\bm{q}'$ are the initial, intermediate, and final relative momenta. $W$ is the energy of the two scattered particles, usually denoted as {\it starting energy}. The Pauli operator $Q$ allows only the scattering to unoccupied states above the Fermi momentum, i.e. $Q(\bm{k},\bm{P})=1$ for $|\bm{k_1}|=|\bm{P}+\bm{k}|> k_F$ and $|\bm{k_2}|=|\bm{P}-\bm{k}|> k_F$ and it vanishes for all other cases.

$E_{\bm{P}\pm\bm{k}}$ are the single-particle energies for nucleons with momenta $\bm{P}\pm\bm{k}$, i.e.
the eigenvalues of the Dirac equation in the nuclear medium:
\beq\label{eq-5}
\left[\bm{\alpha}\cdot\bm{p}+\beta M + \mathcal{U}(\bm{p})\right]\psi({\bm p},s)={\bm E}_{\bm p}\psi({\bm p},s).
\eeq
Here $\psi({\bm p},s)$ is a Dirac-spinor and $\mathcal{U}$ is the relativistic single particle potential
\beq\label{eq-6}
  \langle a|\mathcal{U}|b\rangle =\sum_{c} \langle ac|\hat{G}(W)|bc\rangle.
\eeq
where $|a\rangle$, $|b\rangle$, and $|c\rangle$ are eigenstates of the Dirac operator (\ref{eq-5}) and the index $c$ runs over all the states in the Fermi sea, in analogy to the no-sea approximation of Ref.~\cite{Walecka1974_APNY83-491}.

Due to symmetries, the single-particle potential $\mathcal{U}$ in the rest frame of nuclear matter can be decomposed as
\beq\label{eq-7}
  \mathcal{U}(\bm{p}) = \beta U_S(p)+ U_0(p) + \bm{\alpha\cdot\hat{p}}U_V(p),
\eeq
with the scalar potential $U_S$ and the Lorentz vector potentials $(U_0,\bm{\hat{p}}U_V)$.

As in the non-relativistic case, we have a self-consistent problem that is solved by iteration. The relativistic Hartree-Fock equations are based on an effective (density-dependent) two-body interaction $\hat{G}$ in Eq.(\ref{eq-5}) which is given by the solution of the Thompson equation (\ref{eq-4}) depending in turn on the eigenvalues $E_{\bm{p}}$ of the Dirac operator (\ref{eq-5}).

Besides the problem of the starting energy $W$ in Eq. (\ref{eq-6}), which occurs also in the non-relativistic case (see for instance Refs. \cite{SHEN-SH2018_PLB781-227,SHEN-SH2019_PPNP109-103713}), there are eigenstates of the Dirac operator in Eq. (\ref{eq-5}) with positive energies (PES) and those with negative energies (NES) and it is evident that one needs for the solution of the relativistic Hartree-Fock problem in Eqs. (\ref{eq-5}) and (\ref{eq-6}), in each step of the iteration, not only the matrix elements $\mathcal{U}^{++}$ of the potential (\ref{eq-6}) with positive energies, but also the matrix elements $\mathcal{U}^{+-}$ and $\mathcal{U}^{--}$ with negative energies, i.e. the matrix elements of the $G$-matrix in the full Dirac space. This means, one has to solve the Thompson equation (\ref{eq-4}) in full Dirac space too.

The computer code for the solution of the Thompson equation is rather complicated. It has been developed in the helicity scheme \cite{Erkelenz1971_NPA176-413,Erkelenz1974_PR13-191} for the relativistic scattering process of particles with positive energy to determine the relativistic matrix elements of the bare nucleon-nucleon potential from the scattering phase shifts~\cite{Machleidt1989_ANP19-189}, and it has been used in the literature also in the medium for the calculation of the $G$-matrix elements for states with positive energy: $G^{++++}$. The solution of Eq.~(\ref{eq-4}) leads immediately to the matrix elements $\mathcal{U}^{++}$ of the Dirac potential. In most of the applications of RBHF theory, the remaining matrix elements $\mathcal{U}^{+-}$ and $\mathcal{U}^{--}$ have been derived in various approximations. As examples we mention the Green's function method of Weigel and collaborators~\cite{Poschenrieder1988_PRC38-471,Huber1995_PRC51-1790}, the momentum-independence approximation~\cite{Brockmann1990_PRC42-1965} and the projection methods~\cite{Horowitz1987_NPA464-613,Gross-Boelting1999_NPA648-105}. Solutions including PESs and NESs have been discussed in Refs.~\cite{deJong1998_PRC58-890,Katayama2014_arXiv1410.7166,Katayama2015_PLB747-43}

Today we report on the full solution of the Thompson equation (\ref{eq-4}) in full Dirac space for symmetric nuclear matter using the relativistic potential Bonn A~\cite{Machleidt1989_ANP19-189}. This leads to coupled integral equations for the various channels $G^{++++}$, $G^{+-++}$, $G^{+-+-}$, etc. Details can be found in Ref.~\cite{WANG-SB2021_PRC103-054319}. An additional problem occurs through the fact that the Thompson equation (\ref{eq-4}) is usually solved in the c.m. frame of the scattering process. On the other side the RHF-problem in Eqs.(\ref{eq-5}) and (\ref{eq-6}) is solved in the rest frame of nuclear matter. To avoid this problem, we also solved the Thompson equation in the rest frame of nuclear matter using approximations discussed in detail in Ref.~\cite{WANG-SB2021_PRC103-054319}. In our calculations, we used for the starting energy $W$ the procedure discussed in Ref.~\cite{SHEN-SH2018_PLB781-227}, which means that we consider the matrix elements for negative energy states in Eq. (\ref{eq-6}) in the same way as for occupied states.

\section{Results}
\label{sec-2}
\begin{figure}[h]
\centering
\includegraphics[width=5cm,clip]{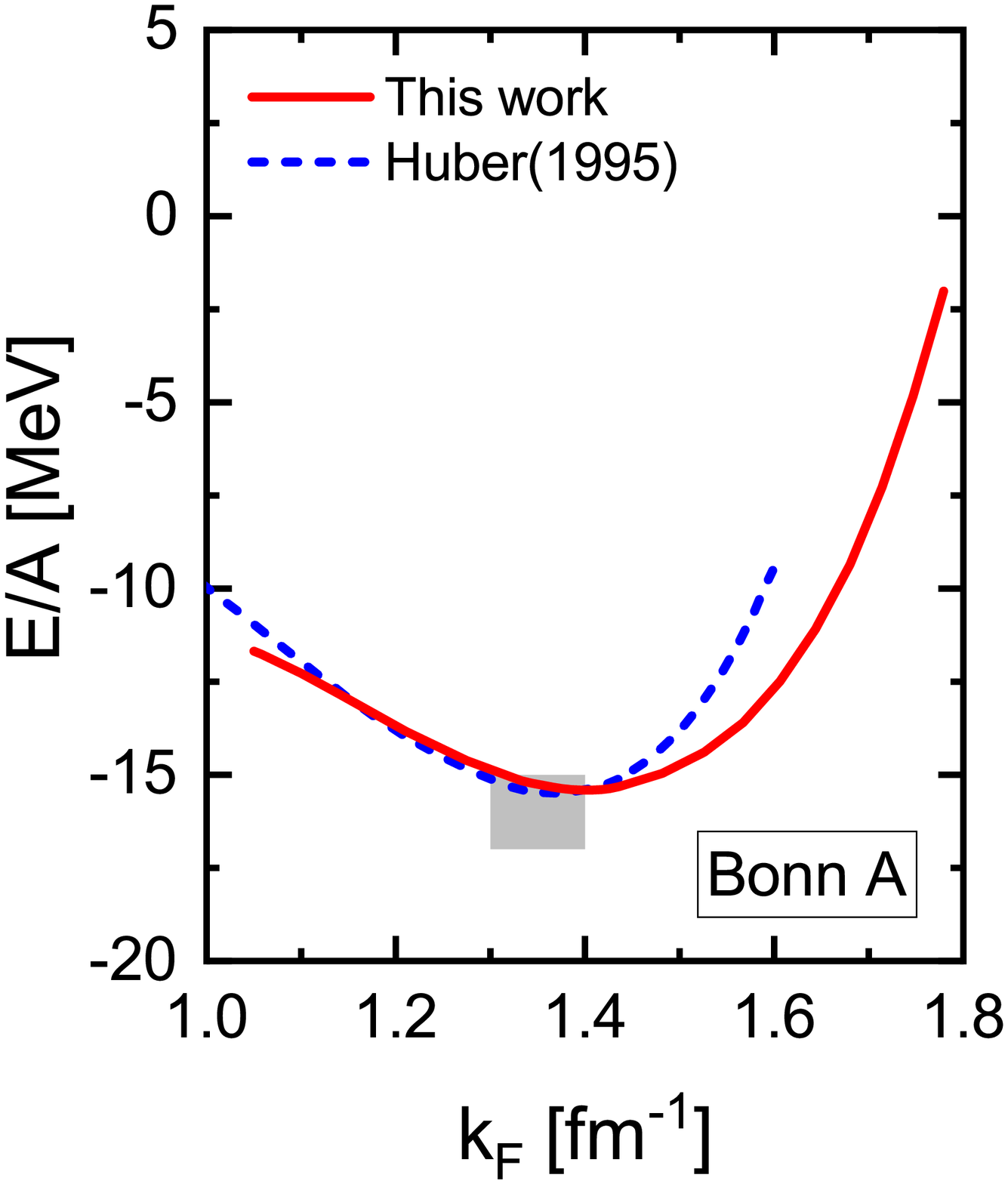}
\hspace{2mm}
\includegraphics[width=5.6cm,clip]{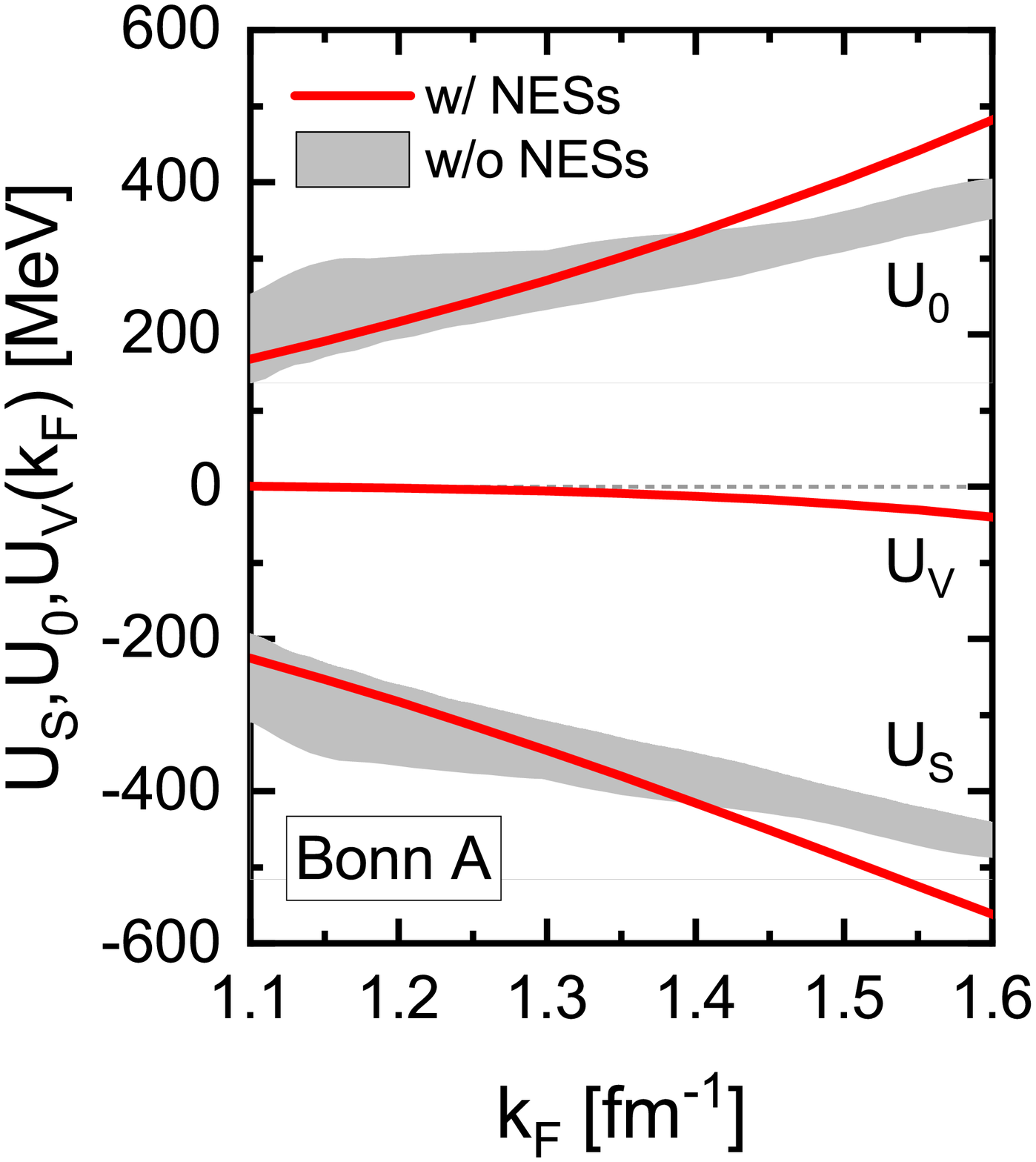}
\caption{(Left side) Binding energy per nucleon $E/A$ in symmetric nuclear matter as a function of the Fermi momentum $k_F$. Treatment with NESs (red) is compared with an approximate treatment of
Ref.~\cite{Huber1995_PRC51-1790}(dashed blue). (Right side) The different contributions to the single particle potential in Eq. (\ref{eq-7}). Results in full Dirac space (red) are compared with those, where the NES are treated approximately (shaded gray). Figures taken from Ref.~\cite{WANG-SB2021_PRC103-054319}}.
\label{fig-1}       
\end{figure}

In the left panel of Fig.~\ref{fig-1}, we show the binding energy per nucleon $E/A$ for symmetric nuclear matter as a function of the Fermi momentum $k_F$. The red line is obtained by RBHF calculations in the full Dirac space with the relativistic potential Bonn A~\cite{Machleidt1989_ANP19-189}. The shaded area indicates the empirical saturation region of symmetric nuclear matter. We find that the nuclear matter saturation point is reasonably described, much better than in a non-relativistic calculation with Bonn A~\cite{Brockmann1990_PRC42-1965}, where the saturation energy is 23.55 MeV and the saturation density corresponds to $k_F=1.85\ \text{fm}^{-1}$ (see Table \ref{tab-1}). The blue line shows the results obtained by the relativistic Green's function approach in the full Dirac space~\cite{Huber1995_PRC51-1790}. They agree with our results below the saturation density. The discrepancy above the saturation density is found mainly arising from the different schemes for the starting energies $W$ in Eq.~\ref{eq-6}: in this work the NESs are treated as occupied, whereas in Ref.~\cite{Huber1995_PRC51-1790} they were treated as unoccupied states.

The right panel of Fig.~\ref{fig-1} shows the different contributions of the relativistic potential in Eq. (\ref{eq-7}) as a function of the Fermi momentum $k_F$. The calculations in the full Dirac space (red) are compared with
results of the momentum-independence approximation of Ref.~\cite{Brockmann1990_PRC42-1965}. In this approximation, $U_V$ is neglected, and the momentum-independent potentials $U_S$ and $U_0$ are extracted from the values of the
single-particle potential energy at two different momenta. The results of this procedure depend on the values
of these two momenta and the corresponding uncertainties are indicated by the shaded areas in the right panel of Fig.~\ref{fig-1}. Although these uncertainties are reduced with the increasing density, the strengths of $U_S$ and $U_0$ are both underestimated above the saturation density, as compared to the results obtained in the full Dirac space. As it is known that the total binding energy depends on the strong cancelation of these two quantities, it is important to solve the RBHF problem in the full Dirac space.

\begin{table}
\centering
\caption{Saturation properties of symmetric nuclear matter. See text for details.}
\label{tab-1}       
\begin{tabular}{lcccc}
\hline
Potential  &   $\rho_0\ [\text{fm}^{-3}]$ &  $E/A\ [\text{MeV}]$ & $K_{\infty}\ [\text{MeV}]$ & $M_D^*/M$ \\ [0pt]
\hline
RBHF Bonn A  &  0.188   &  -15.40 &  258 & 0.55 \\ [-4pt]
RBHF Bonn B  &  0.164   &  -13.36 &  206 & 0.61 \\ [-4pt]
RBHF Bonn C  &  0.144   &  -12.09 &  150 & 0.65 \\ [-4pt]
BHF Bonn A   &  0.428   &  -23.55 &  204 & \\ [-4pt]
BHF Bonn B   &  0.309   &  -18.30 &  160 & \\ [-4pt]
BHF Bonn C   &  0.247   &  -15.75 &  103 & \\ [-4pt]
NL3          &  0.148   &  -16.30 &  272 & 0.60 \\ [-4pt]
DD-ME2       &  0.152   &  -16.14 &  251 & 0.57 \\ [-4pt]
DD-PC1       &  0.152   &  -16.06 &  230 & 0.58 \\ [-4pt]
PC-PK1       &  0.154   &  -16.12 &  238 & 0.59 \\ [-4pt]
PKO1         &  0.152   &  -16.00 &  250 & 0.59 \\ [-4pt]
Empirical    &$\qquad$0.16$\pm0.01$  &  -16$\pm1$ &~~~~~~240$\pm20$ &   \\
\hline
\end{tabular}
\end{table}

In Table \ref{tab-1} we show the saturation properties of symmetric nuclear matter calculated by RBHF theory in the full Dirac space using the interactions Bonn A, B and C~\cite{Machleidt1989_ANP19-189}:
the saturation density $\rho_0$, the binding energy per nucleon $E/A$, the compression modulus $K_\infty$, and the Dirac mass $M_D^*/M$ at saturation density. These quantities are compared with the corresponding values from the non-relativistic BHF calculations and the phenomenological covariant density functionals
NL3~\cite{Lalazissis1997_PRC55-540}, DD-ME2~\cite{Lalazissis2005_PRC71-024312}, DD-PC1~\cite{Niksic2008_PRC78-034318}, PC-PK1~\cite{ZHAO-PW2010_PRC82-054319} and PKO1~\cite{Long2006_PLB640-150} and empirical values for $\rho_0$, $E/A$~\cite{Brockmann1990_PRC42-1965}.

\begin{figure}[h]
\centering
\includegraphics[width=5.8cm,clip]{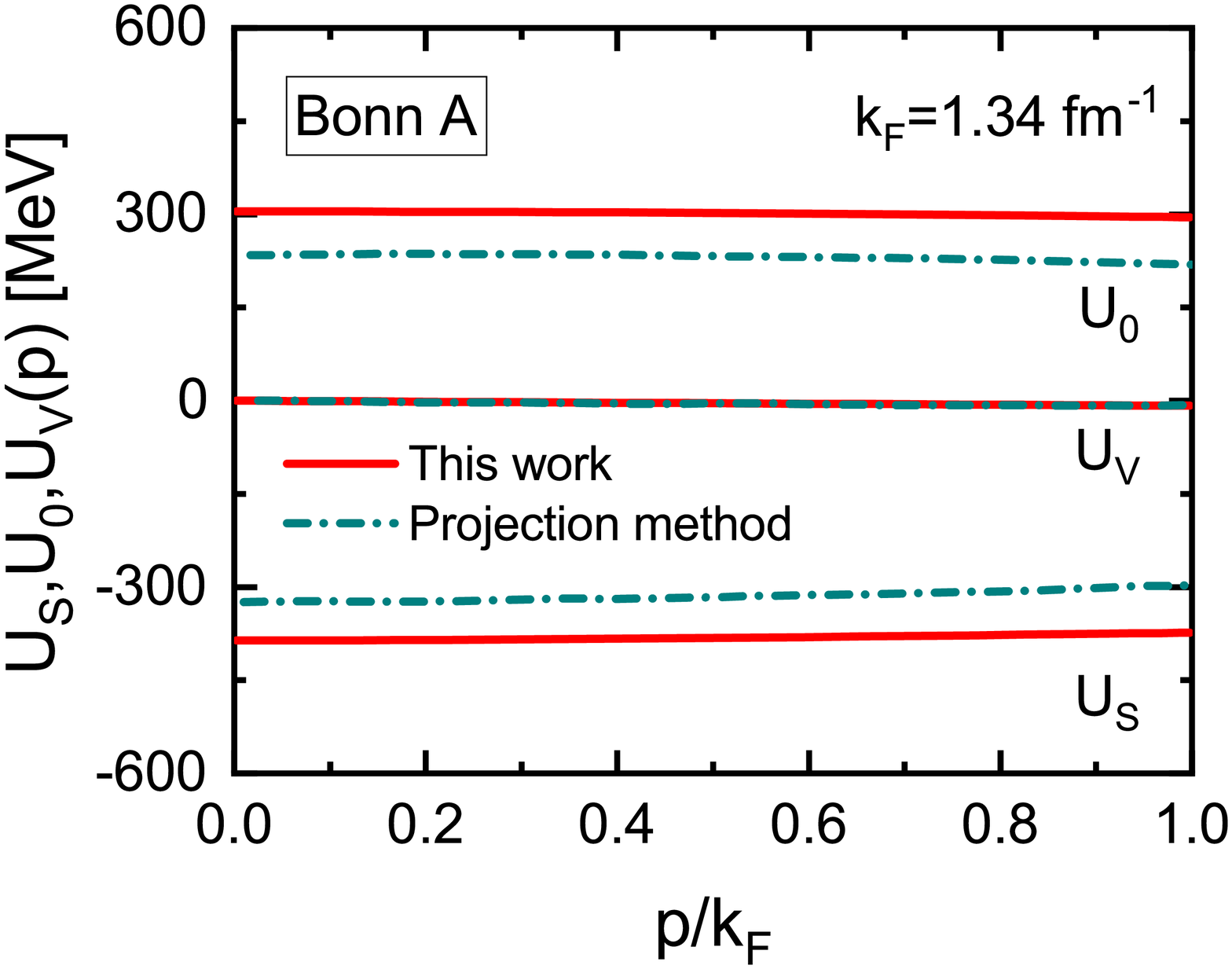}
\hspace{2mm}
\includegraphics[width=5.8cm,clip]{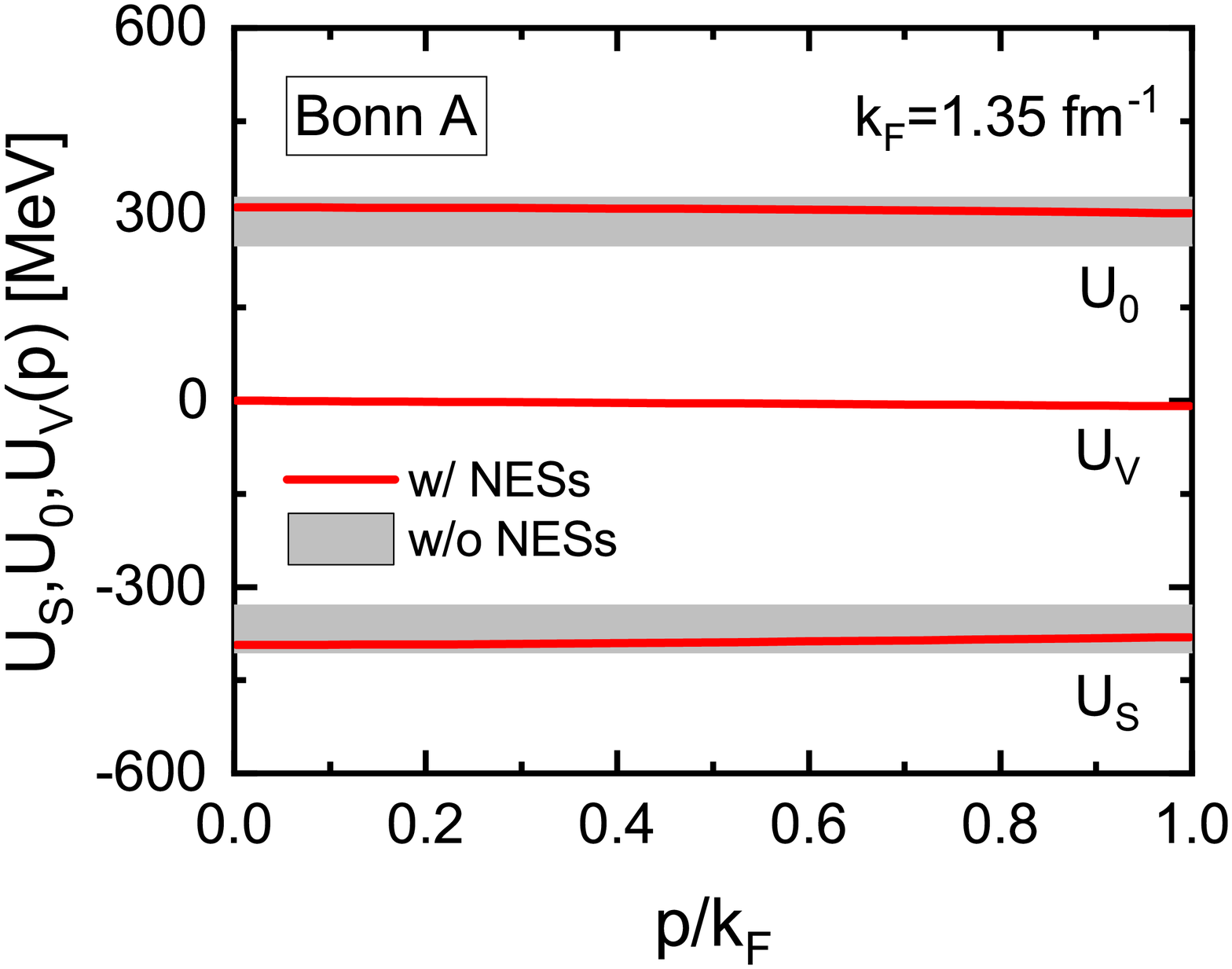}
\caption{Momentum dependence of the single particle potential. Treatment with NESs (red) is compared with two approximations: the projection method of Ref.~\cite{Gross-Boelting1999_NPA648-105} (left side) and the momentum-independence approximation of Ref.~\cite{Brockmann1990_PRC42-1965} (right side). The uncertainties of this approximation are shown by the gray shaded area. Figures taken from Ref.~\cite{WANG-SB2021_PRC103-054319}.}
\label{fig-2}       
\end{figure}

In Fig.~\ref{fig-2}, we show the single-particle potentials $U_S(p)$, $U_0(p)$, and $U_V(p)$ of Eq. (\ref{eq-7}) at saturation density $k_F=1.34$ fm$^{-3}$ as a function of the momentum $p$. The momentum-independence approximation seems to be well justified. The space-like component $U_V(p)$ is also found to be rather small. In the left panel, the calculations in full Dirac space (red) are compared with the projection method of Ref.~\cite{Gross-Boelting1999_NPA648-105} (blue dashed-dotted).
In this method, the $G$ matrix elements are projected onto a complete set of five Lorentz invariant amplitudes~\cite{Horowitz1987_NPA464-613}, and the single-particle potentials are calculated
analytically. However, the choice of these Lorentz invariant amplitudes is not unique. Different schemes of projections have been used~\cite{Horowitz1987_NPA464-613,Gross-Boelting1999_NPA648-105,SHEN-L1997_PRC56-216,Fuchs1998_PRC58-2022} which differ mainly in the effect of the pseudoscalar meson exchange. In the right panel the calculations in full Dirac space (red) are compared with the momentum-independence approximation and its uncertainties (gray shaded areas).

\section{Conclusions}
\label{sec-3}

The RBHF equations have been solved for symmetric nuclear matter in the full Dirac space for potential Bonn A. In this way, one avoids all the uncertainties of previous RBHF calculations without the negative energy states. The Thompson equation is chosen as the scattering equation, and the matrix elements of the Bonn potential are treated in the c.m. frame. The saturation properties of symmetric nuclear matter obtained in this way are in good agreement with the
empirical values. They agree with the results based on the relativistic Green's function approach up to the
saturation density. The discrepancy above the saturation density is found mainly arising from the different schemes for the starting energies. Since there is no clear-cut prescription for the starting energy in the Brueckner-Hartree-Fock method, one has to go beyond the Brueckner approximation for such cases, i.e., one has to include more than two hole lines in the approximation.

An analysis of the uncertainties of the RBHF calculation in the Dirac space with PESs only shows that these uncertainties can reach more than 100 MeV for the single-particle potentials. It also turns out that the equation of state is less bound above the saturation density. These analyses demonstrate the significance of the RBHF calculations in the full Dirac space. Therefore further investigations for asymmetric nuclear matter with the RBHF calculations in the full Dirac space are necessary to clarify the isospin dependence of the single-particle potentials.

\section{Acknowledgements}
This work was supported in part by the National Key R\&D Program of China (No. 2017YFE0116700 and No. 2018YFA0404400), the National Natural Science Foundation of China (No. 11935003, No. 11975031, No. 11875075, and No. 12070131001), and
the Deutsche Forschungsgemeinschaft (DFG, German Research Foundation) under Germany's Excellence Strategy
EXC-2094-390783311, ORIGINS. Part of this work was achieved by using the High-performance Computing Platform of Peking University, and the supercomputer OCTOPUS at the Cybermedia Center, Osaka University under the support of Research Center for Nuclear Physics of Osaka University.


\end{document}